\documentclass[english,letterpaper,aps,prl,twocolumn,showpacs]{revtex4-1}
\usepackage{lmodern}

\usepackage[T1]{fontenc}
\usepackage[latin1]{inputenc}
\usepackage{graphicx}
\usepackage{amssymb}

\makeatletter

\usepackage{lmodern}
\usepackage{subfig}

\makeatletter

\usepackage{lmodern}

\makeatletter

\usepackage{lmodern}

\makeatletter
\usepackage{lmodern}

\makeatletter

\makeatletter

\usepackage{epsfig}

\usepackage{dcolumn}

\usepackage{bm}

\usepackage{bbm}

\usepackage{wasysym}

\newlength{\textwidthm}

\makeatother

\makeatother

\makeatother

\makeatother

\makeatother

\usepackage{babel}
\makeatother
\begin{document}

\title{Why  Dirac points in graphene are where they are?}
%\shorttitle{} %Insert here a short version of the title if it exceeds 70 characters

\author{E. Kogan}
\email{Eugene.Kogan@biu.ac.il}

\affiliation{Department of Physics, Bar-Ilan University, Ramat-Gan 52900,
Israel}
\date{\today}

\begin{abstract}
We present a simple group theory explanation of the fact that the energy bands merge  in the corners of the Brillouin zone  for   graphene and for two particular cases of Kagome lattice
for arbitrary tight--binding Hamiltonian. We connect the linearity of the spectrum in the vicinity of these points for monolayer graphene, bilayer graphene for AA stacking and  Kagome lattice with the properties of the conical points of the surface, known from  geometry.

\end{abstract}


\maketitle

\section{Introduction}

The contacts of energy bands have been studied from the early days of quantum solid state physics \cite{hund,herring}.
The early classical papers in the field are reprinted in Ref. \cite{knox}.
The existence of  such points of contact in graphene was also  known since the early quantum mechanical studies of graphite. When graphene was first isolated experimentally \cite{novoselov}, these points were immediately brought into
the focus of attention of both   theorists and experimentalists. The reason is very simple: for undoped and ungated graphene these points lie on the Fermi surface,
and many physical properties of such graphene are determined by the electron states in the vicinity of these points. In addition, the electron  spectrum
in the vicinity of these points is linear (in a monolayer graphene), and these states are effectively described by the Dirac equation \cite{castro}. This is why these points
 of contact in monolayer graphene (and other materials showing similar properties)  are called the Dirac points.

Graphene is traditionally considered within the tight--binding approximation. The approximation starts from mentioning that
in isolated form, carbon has six electrons in the orbital configuration
$1s^{2}2s^{2}2p^{2}$. When arranged in the honeycomb crystal, two electrons remain in the core
$1s$ orbital, while the other orbitals hybridize, forming three $sp^{2}$
bonds and one $p_{z}$ orbital. The $sp^{2}$ orbitals form the $\sigma$
band, which contains three localized electrons. The bonding configuration
among the $p_{z}$ orbitals of different lattice sites generates a
valence band, or $\pi$-band, containing one electron, whereas the
antibonding configuration generates the conduction band ($\pi^{*}$),
which is empty \cite{castro}.

The early classical papers consider merging of Brillouin zones \cite{knox}.
On the other hand,  in the framework of the tight-binding approximation there exists only a single Brillouin zone. However, this zone contains the number of energy bands, equal to the number of atoms in the elementary cell (more exactly, the number of orbitals per elementary cell considered), which emulate a finite number of Brillouin zones.

The structure of graphene
can be seen as a triangular lattice with a basis of two
atoms per unit cell.
For a model tight--binding Hamiltonian, taking into account only nearest--neighbor  and next--nearest--neighbor hopping, it can be easily shown that the energy bands
these bands neither overlap, nor have a finite  gap between them, but touch each other in two points
in momentum space, both for  monolayer and bilayer graphene.
  \cite{castro}.  We
 show in the framework of group theory that  taking into account hopping between more distant neighbors does not shift these points of contact; their
 positions are determined only by the point symmetry of the honeycomb lattice. The linearity of the spectrum of the electrons
 in the vicinity of
 the points of contact in monolayer and bilayer graphene with the AA stacking also remains for the general  tight--binding Hamiltonian.

In addition we prove the existence of the Dirac points in the corners of the bands for the isotropic  Kagome and Lieb lattices, described in the framework of the general tight--binding Hamiltonian.

\section{The general tight--binding Hamiltonian}

The structure of graphene
can be seen as a triangular lattice with a basis of two
atoms per unit cell, displaced from each other by any one (fixed) vector connecting two sites of different sub-lattices, say
${\bf \delta}=-a\left(1,0\right)$.
The general  Hamiltonian for the $\pi$ bands is
\begin{eqnarray}
\label{ham}
H =-
\left(\begin{array}{cc}
\sum_{\bf a} t'({\bf a})e^{i{\bf k\cdot a}} & \sum_{\bf a}t({\bf a}+{\bf \delta})e^{i{\bf k\cdot}({\bf a}+{\bf \delta})}\\
\sum_{\bf a}t^*({\bf a}+{\bf \delta})e^{-i{\bf k\cdot}({\bf a}+{\bf \delta})} &  \sum_{\bf a}t'({\bf a})e^{i{\bf k\cdot a}} \end{array}\right),\nonumber\\
\end{eqnarray}
where ${\bf a}$ is an arbitrary  lattice vector, that is a linear combination of
${\bf a}_1=\frac{a}{2}\left(3,\sqrt{3}\right)$, ${\bf a}_2=\frac{a}{2}\left(3,-\sqrt{3}\right)$.

The selection rule for matrix elements
\cite{landau} gives
\begin{eqnarray}
\label{zero}
\sum_{\bf a}t({\bf a}+{\bf \delta})e^{i{\bf K\cdot}({\bf a}+{\bf \delta})}=0,
\end{eqnarray}
where {\bf K} is a the corner of the Brillouin zone.
In fact, we are dealing with the product of two functions: $t({\bf a}+{\bf \delta})$ realizes the unit representation
of the point symmetry group $C_{3}$ (the full symmetry group of the inter--sublattice hopping is $C_{3v}$, but the restricted symmetry $C_3$  will be enough to prove the cancelation).
As far as the function $e^{i{\bf K\cdot}({\bf a}+{\bf \delta})}$ is concerned, rotation of the lattice by the angle $2\pi/3$,
say anticlockwise,
 is equivalent to rotation of the vector ${\bf K}$ in the opposite direction,
 that is to substitution of the three equivalent corners of the Brilluoin zone: ${\bf K}_1\to {\bf K}_2\to{\bf K}_3\to {\bf K}_1$, where
${\bf K}_1=\left(\frac{2\pi}{3a},\frac{2\pi}{3\sqrt{3}a}\right)$, ${\bf K}_2=\left(0,-\frac{4\pi}{3\sqrt{3}a}\right)$
and ${\bf K}_3=\left(-\frac{2\pi}{3a},\frac{2\pi}{3\sqrt{3}a}\right)$. Thus due to the rotation $e^{i{\bf K\cdot}({\bf a}+{\bf \delta})}$
is multiplied by the factor $\epsilon^2$ and
realizes  $x-iy$ representation of the group $C_{3}$. Because each of multipliers in Eq. (\ref{zero}) realizes different irreducible representation of the symmetry group, the matrix element is equal to zero.
Simply speaking, at a point {\bf K} the sublattices become decoupled, and this explains the degeneracy of the electron states in this point (these points) or, in other words, merging of the two branches of the single Brilouin zone.

On the other hand, generally
\begin{eqnarray}
\sum_{\bf a} t'({\bf a})e^{i{\bf K\cdot a}}\neq 0.
\end{eqnarray}
To understand this statement  consider the maximum symmetry group of the intra--lattice hopping: $C_{6v}$.
The function $t'({\bf a})$ realizes the $A_1$ representation
of the  group.
Applying  Eq. (\ref{ex}) we see that reducible representation   of the group $C_{6v}$,  realized by the two functions $e^{i{\bf K\cdot a}}$ and $e^{i{\bf K'\cdot a}}$ can be decomposed as $A_1+B_2$.

In addition, the tight--binding model provides us with a simple explanation why the dispersion law in the vicinity of the merging points is linear, that is why these points are Dirac points.
The dispersion law for the Hamiltonian (\ref{ham}) is given by equation
\begin{eqnarray}
\label{dispersion}
F(E,{\bf k})=0,
\end{eqnarray}
where
\begin{eqnarray}
\label{dispersion2}
F(E,{\bf k})=\left|\begin{array}{cc}
E+\sum_{\bf a} t'({\bf a})e^{i{\bf k\cdot a}} & \sum_{\bf a}t({\bf a}+{\bf \delta})e^{i{\bf k\cdot}({\bf a}+{\bf \delta})}\\
\sum_{\bf a}t^*({\bf a}+{\bf \delta})e^{-i{\bf k\cdot}({\bf a}+{\bf \delta})} & E+ \sum_{\bf a}t'({\bf a})e^{i{\bf k\cdot a}} \end{array}\right|.\nonumber\\
\end{eqnarray}

In mathematics
the Dirac points, we are dealing with, are called conical points of the surface; if the surface is given by Eq. (\ref{dispersion}),
the conditions for the conical points are \cite{goursat}
\begin{eqnarray}
\label{dirac}
\frac{\partial F}{\partial E}=0,\qquad
 \frac{\partial F}{\partial {\bf k}}=0.
\end{eqnarray}
Recalling the rule for  differentiating of a determinant, we realize that Eq. (\ref{zero}) guaranties that the conditions (\ref{dirac}) for ${\bf k}={\bf K}$.
This explains linearity of the spectrum in the vicinity of the points {\bf K}({\bf K}').
The  axis of the cone is perpendicular to the $k_x,k_y$ plane, and the cone is a circular one,
because  any vector in the $k_x,k_y$ plane compatible with the symmetry $C_{3v}$ is identically equal to zero, and any  tensor of rank two
 compatible with the symmetry  is proportional to the unity tensor.

The important role played by discrete symmetries in protecting
a $k$--linear dispersion in graphene was pointed out  in
Ref. \cite{manes}.
The appearance of massless Dirac fermions  under conditions of hexagonal symmetry was considered in Ref. \cite{park}.
Group theory was
used to derive an invariant expansion of the Hamiltonian for electron states near the K points of the graphene
Brillouin zone  in Ref. \cite{winkler}.

\section{Bilayer graphene}

The same selection rule for matrix elements, we applied for the case of monolayer graphene, can be applied to bilayer graphene.
\begin{figure}[h]
\begin{center}
\centering
\includegraphics[width=0.2\textwidth]{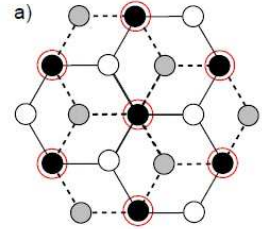}
%\vskip -4cm
\caption{\label{fig:bilayer}Top view of a graphene bilayer; white and black circles: top layer carbon atoms; gray and red: bottom layer. [Adapted from Ref. \cite{uchoa}]}
\end{center}
\end{figure}

 The general tight--binding  Hamiltonian for the AB stacking  has the form
\begin{widetext}
\begin{eqnarray}
\label{ham2}
H =-
\left(\begin{array}{cccc}
\sum_{\bf a} t'({\bf a})e^{i{\bf k\cdot a}} & \sum_{\bf a}t({\bf a}+{\bf \delta})e^{i{\bf k\cdot}({\bf a}+{\bf \delta})} &
\sum_{\bf a}t_4({\bf a}+{\bf \delta})e^{i{\bf k\cdot}({\bf a}+{\bf \delta})} & \sum_{\bf a}t_3({\bf a}+{\bf \delta})e^{i{\bf k\cdot}({\bf a}+{\bf \delta})}\\
\sum_{\bf a}t^*({\bf a}+{\bf \delta})e^{-i{\bf k\cdot}({\bf a}+{\bf \delta})} &  \sum_{\bf a}t'({\bf a})e^{i{\bf k\cdot a}} &
 \sum_{\bf a} t_1({\bf a})e^{i{\bf k\cdot a}} & \sum_{\bf a}t_4({\bf a}+{\bf \delta})e^{i{\bf k\cdot}({\bf a}+{\bf \delta})} \\
\sum_{\bf a}t_4^*({\bf a}+{\bf \delta})e^{-i{\bf k\cdot}({\bf a}+{\bf \delta})} &  \sum_{\bf a} t_1^*({\bf a})e^{-i{\bf k\cdot a}} & \sum_{\bf a} t'({\bf a})e^{i{\bf k\cdot a}} & \sum_{\bf a}t({\bf a}+{\bf \delta})e^{i{\bf k\cdot}({\bf a}+{\bf \delta})}\\
 \sum_{\bf a}t_3^*({\bf a}+{\bf \delta})e^{-i{\bf k\cdot}({\bf a}+{\bf \delta})} &
  \sum_{\bf a}t_4^*({\bf a}+{\bf \delta})e^{-i{\bf k\cdot}({\bf a}+{\bf \delta})} &
\sum_{\bf a}t^*({\bf a}+{\bf \delta})e^{-i{\bf k\cdot}({\bf a}+{\bf \delta})}  & \sum_{\bf a} t'({\bf a})e^{i{\bf k\cdot a}}
 \end{array}\right),\nonumber\\
\end{eqnarray}
\end{widetext}
where $t'$ and $t$ are the intra-- and inter--sublattice intra--layer hopping integrals respectively, and
$t_1$, $t_3$ and $t_4$ are the $A-A$, $B-B$, and $A-B$ inter--layer hopping integrals respectively \cite{malard}.

For ${\bf k}={\bf K}({\bf K})'$, the  terms containing $t,t_3$ and $t_4$ are zero, due to symmetry reasons mentioned above. Hence, at the corner of the
Brillouin zone the Hamiltonian (\ref{ham2}) takes the form
\begin{widetext}
\begin{eqnarray}
\label{ham3}
H =-
\left(\begin{array}{cccc}
\sum_{\bf a} t'({\bf a})e^{i{\bf K\cdot a}} & 0 & 0 & 0 \\
0 &  \sum_{\bf a}t'({\bf a})e^{i{\bf K\cdot a}} & \sum_{\bf a} t_1({\bf a})e^{i{\bf K\cdot a}} & 0 \\
0 &  \sum_{\bf a} t_1^*({\bf a})e^{-i{\bf K\cdot a}} & \sum_{\bf a} t'({\bf a})e^{i{\bf K\cdot a}} & 0 \\
0 & 0 & 0  & \sum_{\bf a} t'({\bf a})e^{i{\bf K\cdot a}}
 \end{array}\right).
\end{eqnarray}
\end{widetext}
We again have merging of the energy band. Of course, in this case  the points of contact are not the Dirac points.

The general tight--binding  Hamiltonian for the AA stacking  has the form
\begin{widetext}
\begin{eqnarray}
\label{ham22}
H =-
\left(\begin{array}{cccc}
\sum_{\bf a} t'({\bf a})e^{i{\bf k\cdot a}} & \sum_{\bf a}t({\bf a}+{\bf \delta})e^{i{\bf k\cdot}({\bf a}+{\bf \delta})} &
\sum_{\bf a} t_1({\bf a})e^{i{\bf k\cdot a}} & \sum_{\bf a}t_4({\bf a}+{\bf \delta})e^{i{\bf k\cdot}({\bf a}+{\bf \delta})}\\
\sum_{\bf a}t^*({\bf a}+{\bf \delta})e^{-i{\bf k\cdot}({\bf a}+{\bf \delta})} &  \sum_{\bf a}t'({\bf a})e^{i{\bf k\cdot a}} &
\sum_{\bf a}t_4({\bf a}+{\bf \delta})e^{i{\bf k\cdot}({\bf a}+{\bf \delta})} & \sum_{\bf a} t_1({\bf a})e^{i{\bf k\cdot a}} \\
\sum_{\bf a} t_1({\bf a})e^{i{\bf k\cdot a}} &   \sum_{\bf a}t_4^*({\bf a}+{\bf \delta})e^{-i{\bf k\cdot}({\bf a}+{\bf \delta})} & \sum_{\bf a} t'({\bf a})e^{i{\bf k\cdot a}} & \sum_{\bf a}t({\bf a}+{\bf \delta})e^{i{\bf k\cdot}({\bf a}+{\bf \delta})}\\
 \sum_{\bf a}t_4^*({\bf a}+{\bf \delta})e^{-i{\bf k\cdot}({\bf a}+{\bf \delta})} &
\sum_{\bf a} t_1({\bf a})e^{i{\bf k\cdot a}} &
\sum_{\bf a}t^*({\bf a}+{\bf \delta})e^{-i{\bf k\cdot}({\bf a}+{\bf \delta})}  & \sum_{\bf a} t'({\bf a})e^{i{\bf k\cdot a}}
 \end{array}\right),\nonumber\\
\end{eqnarray}
\end{widetext}
where
$t_1$ and $t_4$ are the $A-A(B-B)$  and $A-B$ inter--layer hopping integrals respectively. At the corner of the
Brillouin zone the Hamiltonian (\ref{ham22}) takes the form
\begin{widetext}
\begin{eqnarray}
\label{ham222}
H =-
\left(\begin{array}{cccc}
\sum_{\bf a} t'({\bf a})e^{i{\bf K\cdot a}} & 0 & \sum_{\bf a} t_1({\bf a})e^{i{\bf K\cdot a}} & 0 \\
0 &  \sum_{\bf a}t'({\bf a})e^{i{\bf K\cdot a}} & 0 & \sum_{\bf a} t_1({\bf a})e^{i{\bf K\cdot a}} \\
\sum_{\bf a} t_1({\bf a})e^{i{\bf K\cdot a}} &  0 & \sum_{\bf a} t'({\bf a})e^{i{\bf K\cdot a}} & 0 \\
0 & \sum_{\bf a} t_1({\bf a})e^{i{\bf K\cdot a}} &
0  & \sum_{\bf a} t'({\bf a})e^{i{\bf K\cdot a}}
 \end{array}\right),\nonumber\\
\end{eqnarray}
\end{widetext}
and we again have merging of two bands. However, for the AA stacking, in distinction from the AB stacking the function
\begin{eqnarray}
F(E,{\bf k}) =\left|EI-H\right|,
\end{eqnarray}
where $I$ is the unity matrix, satisfies Eq. (\ref{dirac}). Hence the points of contact are the Dirac points.

\section{Kagome lattice}

Kagome lattice has three sublattices. The general anisotropic tight--binding Hamiltonian for such lattice is characterized by three sets of exchange integrals.  We consider general tight--binding Hamiltonian for two particular cases.
First consider isotropic case $t_1=t_2=t_3=t$.
The Hamiltonian in this case is given by equation
\begin{widetext}
\begin{eqnarray}
\label{ham5}
H =
\left(\begin{array}{ccc}
\sum_{\bf a} t({\bf a})e^{i{\bf k\cdot a}} & \sum_{\bf a}t\left({\bf a}+{\bf a}_1/2\right)e^{i{\bf k\cdot}\left({\bf a}+{\bf a}_1/2\right)} &
\sum_{\bf a}t\left({\bf a}+{\bf a}_3/2\right)e^{i{\bf k\cdot}\left({\bf a}+{\bf a}_3/2\right)}\\
\sum_{\bf a}t^*\left({\bf a}+{\bf a}_1/2\right)e^{-i{\bf k\cdot}\left({\bf a}+{\bf a}_1/2\right)} & \sum_{\bf a} t({\bf a})e^{i{\bf k\cdot a}} &
\sum_{\bf a}t\left({\bf a}+{\bf a}_2/2\right)e^{i{\bf k\cdot}\left({\bf a}+{\bf a}_2/2\right)}\\
\sum_{\bf a}t^*\left({\bf a}+{\bf a}_3/2\right)e^{-i{\bf k\cdot}\left({\bf a}+{\bf a}_3/2\right)} &
\sum_{\bf a}t^*\left({\bf a}+{\bf a}_2/2\right)e^{-i{\bf k\cdot}\left({\bf a}+{\bf a}_2/2\right)} & \sum_{\bf a} t({\bf a})e^{i{\bf k\cdot a}} \end{array}\right),
\end{eqnarray}
\end{widetext}
where ${\bf a}_1$ and  ${\bf a}_2$ are the elementary lattice vectors, and
${\bf a}_3={\bf a}_1-{\bf a}_2$.

Eq. (\ref{ham5}) can be presented as
%\begin{widetext}
\begin{eqnarray}
\label{ham10}
H =
\left(\begin{array}{ccc}
V({\bf k}) & T({\bf k})\cos\left(\frac{{\bf k\cdot a}_1}{2}\right) & T({\bf k})\cos\left(\frac{{\bf k\cdot a}_3}{2}\right)\\
T({\bf k})\cos\left(\frac{{\bf k\cdot a}_1}{2}\right) & V({\bf k}) & T({\bf k})\cos\left(\frac{{\bf k\cdot a}_2}{2}\right)\\
T({\bf k})\cos\left(\frac{{\bf k\cdot a}_3}{2}\right) & T({\bf k})\cos\left(\frac{{\bf k\cdot a}_2}{2}\right) & V({\bf k}) \end{array}\right),\nonumber\\
\end{eqnarray}
%\end{widetext}
with obvious expressions for $V({\bf k})$ and $T({\bf k})$. We see, that the dispersion law for a general tight--binding Hamiltonian
can be presented as
\begin{eqnarray}
\label{reduced}
E({\bf k})=V({\bf k})+T({\bf k})\tilde{E}({\bf k})
\end{eqnarray}
where $\tilde{E}({\bf k})$ is the dispersion law given by the Hamiltonian  taking into account only the nearest--neighbor hopping
\cite{ziegler}
\begin{eqnarray}
\label{ham11}
\tilde{H} =
\left(\begin{array}{ccc}
0 & \cos\left(\frac{{\bf k\cdot a}_1}{2}\right) & \cos\left(\frac{{\bf k\cdot a}_3}{2}\right)\\
\cos\left(\frac{{\bf k\cdot a}_1}{2}\right) & 0 & \cos\left(\frac{{\bf k\cdot a}_2}{2}\right)\\
\cos\left(\frac{{\bf k\cdot a}_3}{2}\right) &  \cos\left(\frac{{\bf k\cdot a}_2}{2}\right) & 0 \end{array}\right).
\end{eqnarray}
 One branch of the spectrum we find  by inspection:
\begin{eqnarray}
\tilde{E}_1({\bf k})=-1.
\end{eqnarray}
The other two branches are given by the reduced Hamiltonian
\begin{eqnarray}
\label{ham12}
\hat{H} =\frac{1}{2}
\left(\begin{array}{cc}
1 & \sum_{i=1}^3e^{i{\bf k\cdot a}_i}\\
\sum_{i=1}^3e^{-i{\bf k\cdot a}_i} & 1 \end{array}\right).
\end{eqnarray}
At the point K(K') the non--diagonal matrix elements of the Hamiltonian (\ref{ham12}) are equal to zero due to the reasons presented above,
and we have  merging of the bands.
In addition, at the point K(K') the function $F$, defined by
\begin{eqnarray}
F(\tilde{E},{\bf k}) \equiv
\left|\begin{array}{cc}
1-\tilde{E} & \sum_{i=1}^3e^{i{\bf k\cdot a}_i}\\
\sum_{i=1}^3e^{-i{\bf k\cdot a}_i} & 1-\tilde{E} \end{array}\right|,
\end{eqnarray}
satisfies conditions (\ref{dirac}).
Thus we have two Dirac points in the corners of the Brillouin zone. Note also, that
 at ${\bf \Gamma}$ point one of the branches given by Eq. (\ref{ham12}) touches the branch $\tilde{E}_1$.

Now consider particular case of the Kagome lattice $t_1=t_2=t$, $t_3=0$. In this case, called isotropic Lieb lattice,
we  recover Eq. (\ref{reduced}), the role of the Hamiltonian (\ref{ham11}) being played by the Hamiltonian
\begin{eqnarray}
\label{ham16}
\tilde{H} =
\left(\begin{array}{ccc}
0 & \cos\left(\frac{{\bf k\cdot a}_1}{2}\right) & 0 \\
\cos\left(\frac{{\bf k\cdot a}_1}{2}\right) & 0 & \cos\left(\frac{{\bf k\cdot a}_2}{2}\right)\\
0 &  \cos\left(\frac{{\bf k\cdot a}_2}{2}\right) & 0 \end{array}\right).
\end{eqnarray}
The spectrum is
\begin{eqnarray}
\tilde{E}_1({\bf k})=0,\quad \tilde{E}_{2,3}^2({\bf k})=\cos^2\left(\frac{{\bf k\cdot a}_1}{2}\right) +
\cos^2\left(\frac{{\bf k\cdot a}_1}{2}\right).\nonumber\\
\end{eqnarray}
This time all three bands merge at the point ${\bf M}=\left(\frac{2\pi}{3a},0\right)$, the zones $E_2$ and $E_3$ again
forming Dirac cones in the vicinity of the point of contact.

In this work we presented  simple arguments explaining merging of the energy bands in the corners of the Brillouin zone
for the case of both monolayer and bilayer graphene and two particular cases of the Kagome lattice and the Dirac nature of these merging point
in the case of monolayer graphene,  bilayer graphene for AA stacking and Kagome lattice.

I am  grateful to K. Ziegler, V. Nazarov, G.-Yu. Guo and J. L. Manes for very useful discussions.

The work was done during the author's participation in "1st Workshop on Nanoscience: Graphene" in Tainan, Taiwan, December 2011,
and the author's visit to the Academia Sinica and National Chengchi University,  Taipei, Taiwan.

\end{document}